\RequirePackage{lineno}
\documentclass[twocolumn,amsmath,amssymb,10pt,showpacs]{revtex4}
\usepackage{graphicx}
\usepackage{CJK}
\usepackage{dcolumn}
\usepackage{times}
\usepackage{setspace}
\usepackage[pdftex,
          a4paper=true,
          colorlinks,
          citecolor=blue,
          linkcolor=blue,
          urlcolor=blue,
          hyperindex,
          pdfstartview=FitH,
          plainpages=false,
          CJKbookmarks]
          {hyperref}

\begin{document}

\title{Fusion reactions in the $ ^{9} $Be + $ ^{197} $Au system above the Coulomb barrier}

\author{G. S. Li$^{1}$}
\author{J. G. Wang$^{1}$}
\email{wangjg@impcas.ac.cn}
\author{J. Lubian$^{2}$}
\email{lubian@if.uff.br}
\author{H. O. Soler$^{2}$}
\author{Y. D. Fang$^{1}$}
\author{M. L. Liu$^{1}$}
\author{N. T. Zhang$^{1}$}
\author{X. H. Zhou$^{1}$}
\author{Y. H. Zhang$^{1}$}
\author{B. S. Gao$^{1}$}
\author{Y. H. Qiang$^{1}$}
\author{S. Guo$^{1}$}
\author{S. C. Wang$^{1}$}
\author{K. L. Wang$^{1}$}
\author{K. K. Zheng$^{1}$}
\author{R. Li$^{1}$}
\author{Y. Zheng$^{1}$}

\affiliation{%
1~Institute of Modern Physics, Chinese Academy of Sciences, Lanzhou, 73000, People’s Republic of China \\
2~Instituto de F\'{i}sica, Universidade Federal Fluminense, Avenida Litor\^{a}nea s/n, Gragoat\'{a}, Niter\'{o}i, Rio de Janeiro 24210-340, Brazil\\
}%

\date{\today}

\begin{spacing}{1.0}

\begin{abstract}
The cross sections of complete fusion and incomplete fusion for the $ ^{9} $Be + $ ^{197} $Au system, at energies not too much above the Coulomb barrier, were measured for the first time. The online activation followed by an offline $\gamma$-ray spectroscopy method was used for the derivation of the cross sections. A slightly higher value of the incomplete fusion to total fusion ratio has been observed, compared to other systems reported in the literature with a $ ^{9} $Be beam. The experimental data were compared with coupled channel calculations without taking into account the coupling of the breakup channel, and experimental data of other reaction systems with weakly bound projectiles. A complete fusion suppression of about 40\% was found for the $ ^{9} $Be + $ ^{197} $Au system, at energies above the barrier, whereas the total fusion cross sections are in agreement with the calculations.

\end{abstract}
 \maketitle
\end{spacing}

\section{Introduction}\label{sec01}
In recent years, the investigation to the effect  of the breakup of weakly bound projectiles on the fusion cross section has been a subject of intense experimental and theoretical studies \cite{canto2006fusion,keeley2007fusion,liang2005fusion,kolata2016elastic}. Experimentally, the radioactive projectiles such as $^{6}\textrm{He}$\cite{kolata1998sub}, $^{8}\textrm{He}$\cite{lemasson2009modern}, $^{11}\textrm{Li}$\cite{cubero2012halo}, $^{11}\textrm{Be}$\cite{signorini2004subbarrier}, and $^{8}\textrm{B}$\cite{aguilera2011near} have been employed. But due to the low beam intensity, the data statistics are not very high. Therefore, many researchers chose weakly bound stable nuclei $^{6,7}\textrm{Li}$ \cite{dasgupta2002fusion,dasgupta2004effect} and $^{9}\textrm{Be}$ \cite{dasgupta1999fusion,gomes2009near,PhysRevC.91.014608,dasgupta2004effect} for the study. The reason is that the beam intensity of these nuclei can be higher in an order of magnitudes.  Also, the effects of its breakup on other reaction mechanisms, although less intensive, are similar to the ones expected for radioactive beam induced reactions. In the reactions with the weakly bound nuclei, besides the direct complete fusion (DCF) where the projectile fuses with the target, the projectiles may have considerable breakup probability. Following the breakup, different processes may occur: the non-capture breakup (NCBU), when neither fragment fuses with the target; the incomplete fusion (ICF), when part of the fragments fuses; and the sequential complete fusion (SCF), when all the breakup fragments are absorbed sequentially by the target. It is not possible to distinguish between DCF and SCF experimentally, and thus CF is taken as the sum of two processes. Total fusion (TF) is defined as the sum of CF and ICF.
	
A meaningful discussion about fusion with weakly bound nuclei is how the breakup channels affect the fusion cross section, as shown in Refs. \cite{canto2006fusion,keeley2007fusion,liang2005fusion,kolata2016elastic}. To study that, one method widely adopted is performing coupled-channel (CC) calculations without taking into account the breakup and transfer channels. Thus, the differences between the experimental data and theoretical predictions are expected to be due to the coupling effects of the channels not included in the calculations. Some works \cite{Canto_2008,CANTO200951,gomes2009breakup,gomes2011sub,WZG14} have shown that for heavy-ion systems, the breakup and transfer mechanisms of the weakly bound nucleus cause a hindrance in the fusion cross section at energies above the Coulomb barrier and enhancement at energies below it. It was recently proved \cite{GOC12} by means of the calculation of the dynamic polarization potentials (DPP) that the breakup channels produce a repulsive DPP (see also Refs.~\cite{MTE04,LCP07,RLC16,KAR05,SKR11}), and consequently a hindrance of the CF in the whole energy interval around the Coulomb barrier, while the breakup triggered by a transfer of the nucleons produces attractive polarization potentials that enhance the  CF. So, to have enhancement below the barrier, the transfer mechanism predominates, while at energies above the barrier, the direct breakup mechanism should be the most relevant channel that produces the hindrance of the CF cross section \cite{GOC12}. This is in agreement with the experimental results of Refs. \cite{RRR10,LDH11,LDH13}, suggesting that the breakup following transfer is the dominant reaction mechanism at energies below the Coulomb barrier. On the other hand,  recent  experimental  results \cite{CSB19} argued  that the ICF products are compatible with a one-step mechanism, questioning the previous interpretation of a two-step mechanism (breakup followed by the absorption of the fragment).

If one wants to study the systematical feature of fusion cross sections and further plot different fusion excitation functions in the same graphic, it is important that the cross section data should be calculated and transformed in a standard way in which they could be compared to each other, and to a benchmark function. In addition, a proper normalization method should be used for different reaction systems. Canto \textit{et al.} have developed a method, called the universal fusion function (UFF) \cite{Canto_2008,CANTO200951}, which incorporates the above requirements. This method can be applied for different weakly bound systems, with stable and radioactive projectiles. For fusions induced by the weakly bound $^{9}$Be, Gomes \textit{et al.} \cite{gomes2011search} started a systematical study of the CF behavior by comparing the suppression of many different systems, and afterwards more efforts have been devoted to this study \cite{zhang2014complete,PhysRevC.91.014608}, but the investigation was not conclusive. To shed light on the study, this work presents new cross section data of the $^{9}$Be+ $^{197}$Au system, and compares it with other reaction systems using the UFF methodology.

The paper is organized as follows. The experimental setup and measured spectra are presented in Sec.\ref{sec02}. In Sec.\ref{sec03} we present the data reduction and results. Comparison of the data with theoretical calculations and different reaction systems are shown in Sec.\ref{sec04}. The conclusions drawn from the present study are given in Sec.\ref{sec05}.
 	
\section{Experimental details}\label{sec02}

\begin{figure}[!t]
\includegraphics[width=0.45\textwidth]{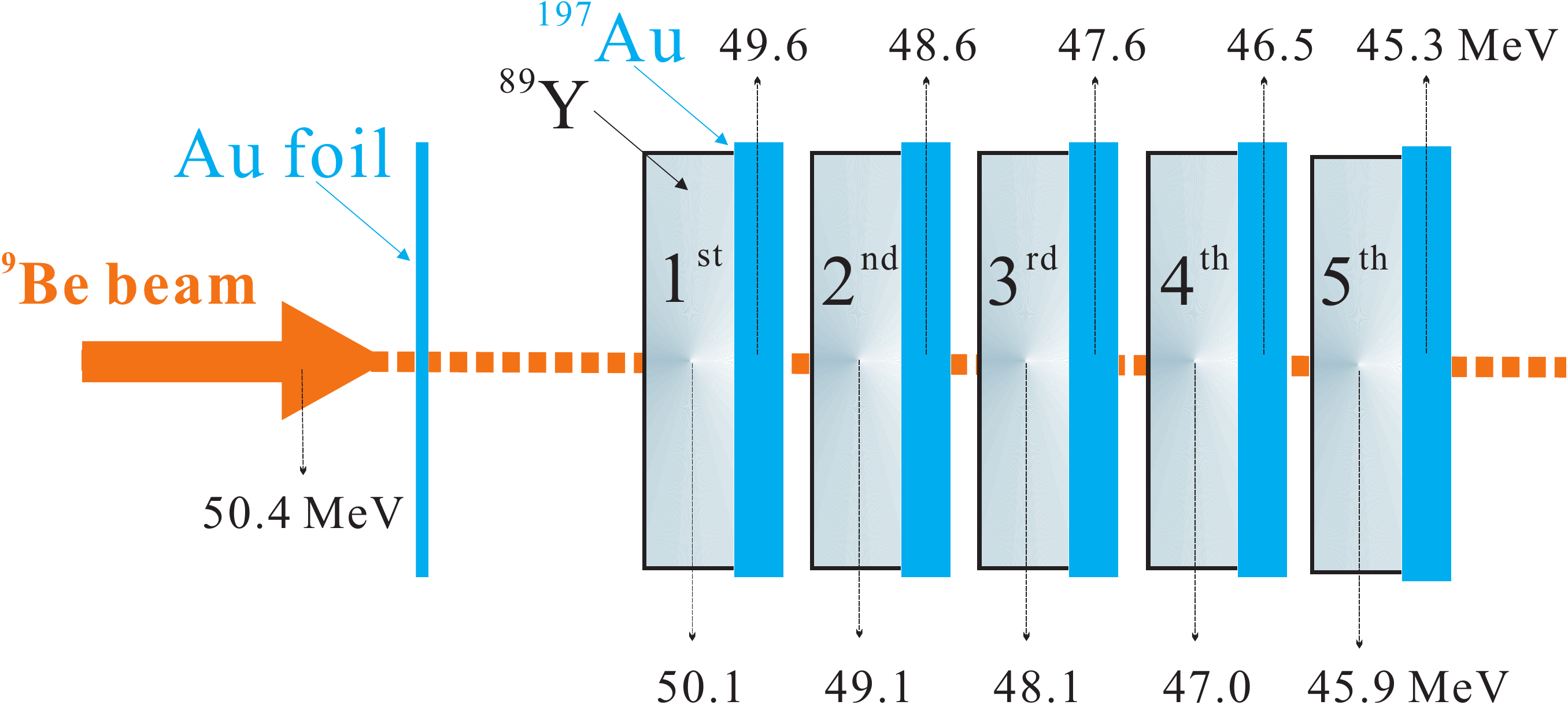}
\caption{\label{fig01}(Color online) Sketch of the beam and target assembly in the experiment. The beam energies in each layer of the assembly is indicated. See text for details.}
\end{figure}

\begin{figure}[!b]
\centering
\includegraphics[width=0.47\textwidth]{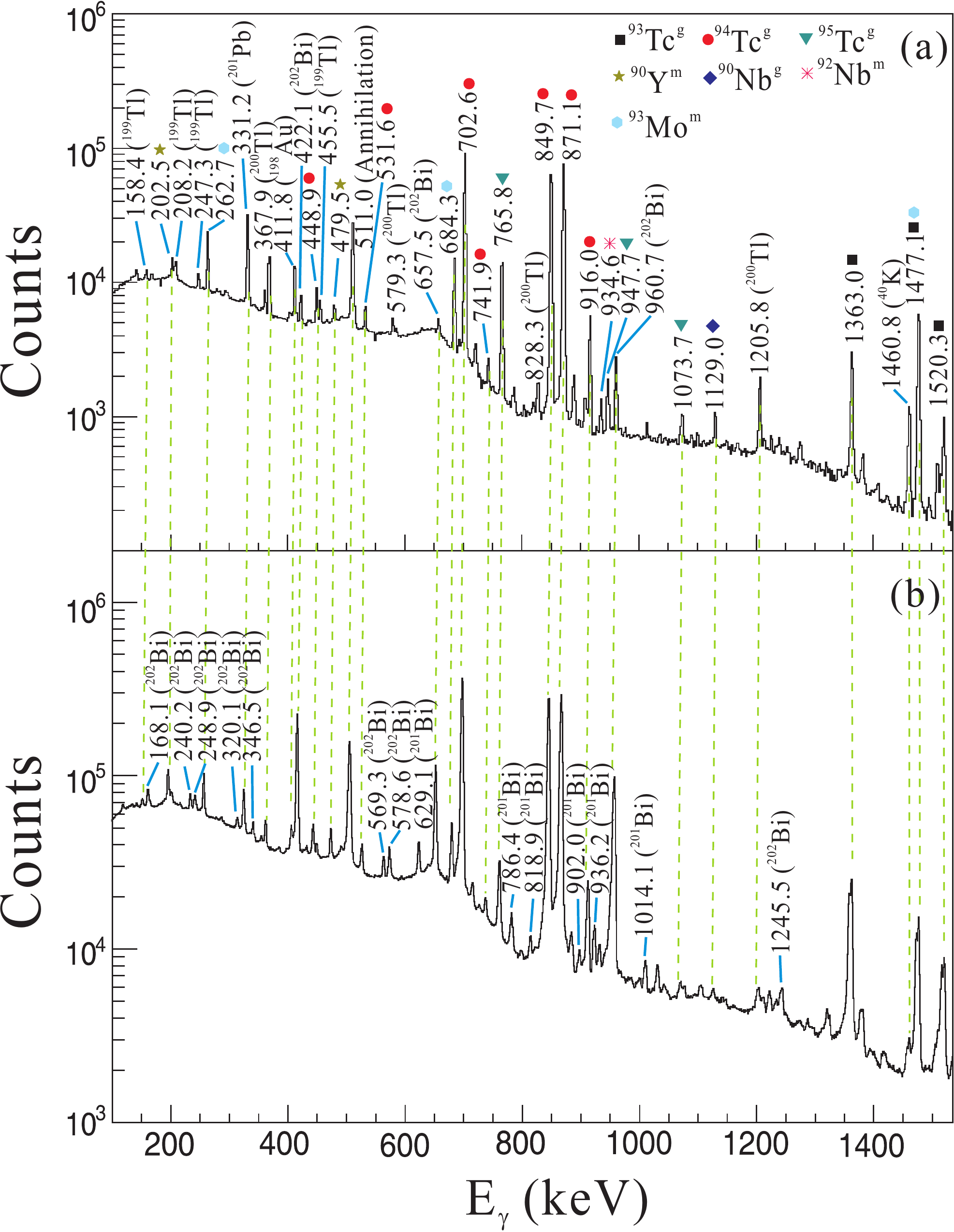}
\caption{\label{fig02} (Color online) Offline $ \gamma $-ray spectra for the $ ^{9} $Be + $ ^{197} $Au system from the first target measured at 10.5 h (a) and 20 min (b) after the end of the activation with a measuring time of 1 h. The contaminated $ \gamma $ rays, mainly from the reaction products of $ ^{9} $Be with $ ^{89} $Y, are indicated. }
\end{figure}
	
The present experiment was carried out through a stacked-foil activation followed by an offline measurement technique. Figure \ref{fig01} shows the sketch of the beam and target assembly for online activation. The collimated $ ^{9} $Be beam with an initial energy of 50.4 MeV was delivered by the Heavy Ion Research Facility in Lanzhou (HIRFL), China. Five targets of $ ^{89} $Y, each with $ ^{197} $Au backing of $ \approx $ 1 mg/cm$ ^{2} $, were irradiated for about 12 h. The average beam current was about 13 enA. The $ ^{197} $Au backing was originally designed as the catcher to trap the recoiling residues produced in the $ ^{9} $Be + $ ^{89} $Y reaction. However, because the $ ^{9} $Be beam could react with the $ ^{197} $Au when passing through the backing, the produced reaction residues were also identified for further analysis. The mean beam energies incident at half the thickness of each $ ^{89} $Y and $ ^{197} $Au backing were indicated in Fig. \ref{fig01}. They were obtained with ATIMA (ATomic Interaction with MAtter) calculation within the LISE++ program \cite{scheidenberger1998penetration,bazin2002program}. The beam flux was calculated by the total charge collected in the Faraday cup placed behind the targets using a high precision current integrator. The Faraday cup was biased with a negative 400 V electrode on the collector to repel the secondary electrons. In addition, during the bombardments, two silicon surface-barrier detectors were placed at $ \pm $30\r{ } to the beam for monitoring the elastic scattering of the beam particles by an Au foil placed upstream from the target stack. In both cases, the profiles of the beam current were recorded by the data acquisition system in intervals of 1 sec. The two sets of flux values were found to agree with each other.

After the irradiation, the activity of the targets was measured offline using five HPGe detector groups in a separate laboratory. Each group consisted of two HPGe detectors positioned 180\r{ } to each other, where single $ \gamma $-ray and $ \gamma $-$ \gamma $ coincidence measurements could be performed simultaneously. The Ge crystal part of each detector was surrounded by a Pb annular cylinder of 3 cm thickness to reduce background from natural radioactivity. In addition, Pb blocks of 6 cm thickness were inserted between each adjacent detector group to shield $ \gamma $ rays from neighboring targets. The absolute efficiency of the detectors was determined using a set of activity calibrated radioactive sources ($ ^{60} $Co, $ ^{133} $Ba, and $ ^{152} $Eu) mounted with the same geometry as the targets. 

\begin{figure}[!bt]
\centering
\includegraphics[width=0.35\textwidth]{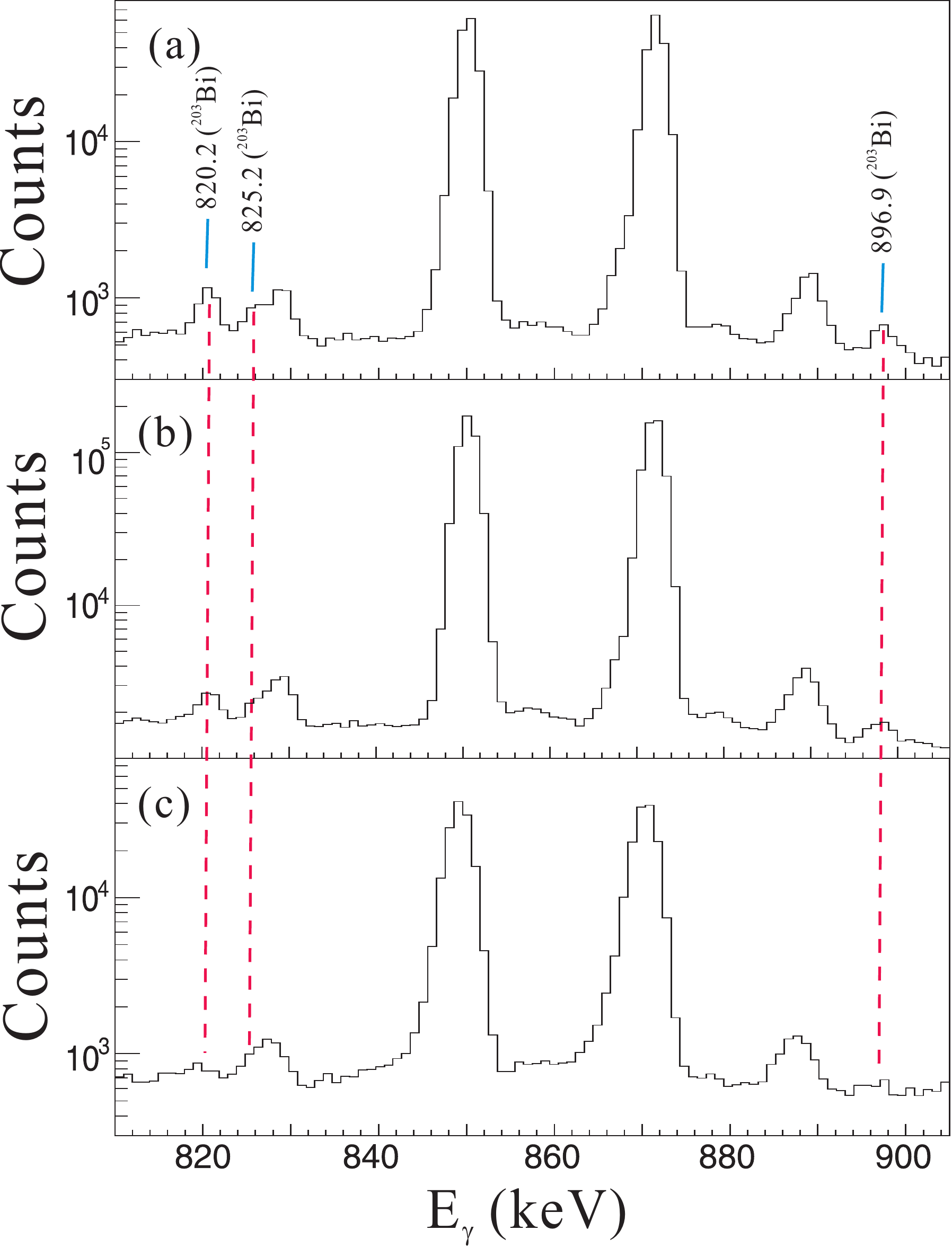}
\caption{\label{fig03} (Color online) Offline $ \gamma $-ray spectra for the $ ^{9} $Be + $ ^{197} $Au system from the first target (a), third target (b), and fifth target (c) measured at 10.5 h after the end of the activation with a measuring time of 1 h. The identification of $ ^{203} $Bi could be justified from the trend of corresponding characteristic $ \gamma $ rays, with changing the beam energies inside different targets. See text for details.}
\end{figure}

\begin{table}[tbp]
\caption{\label{table01}List of evaporation residues identified in the present measurement along with their half-lives $T_{1/2} $, $ J_{\pi} $, $ E_{\gamma} $, and absolute intensities $I _{\gamma} $. The intense $ \gamma $ rays (in bold) were chosen to evaluate the cross sections. The other $ \gamma $ rays corresponding to the same nuclei were also used to cross-check the deduced cross-section values. The decay data was taken from Refs. \cite{KONDEV2007365,ZHU2008699,KONDEV20051,HUANG2016221,SINGH200779,KONDEV20071471}.       }
\begin{ruledtabular}
\setlength{\tabcolsep}{2mm}{
\begin{tabular}{lcccc}
{Residue} &
{$T_{1/2}$} &
{$J^{\pi}$} &
{$E_{\gamma}$(keV)}&
{$I_{\gamma}$(\%)}\\[1mm] \hline
$^{201}$Bi($5n$)  &  103 min  &  9/2$^{-}$  &  \textbf{629.1}  &  26.0\\
                     &         &             &  786.4           &  10.3\\
                     &         &             &  818.9           &  8.0\\  
                     &         &             &  902.0           &  9.0\\ 
                     &         &             &  936.2           &  12.2\\  
                     &         &             &  1014.1          &  11.6\\                  
$^{202}$Bi($4n$)  &  1.71 h  &    5$^{+}$  &  168.1        & 4.8 \\
                     &         &             &    240.2        &  4.5   \\ 
                     &         &             &    248.9        &  3.1   \\ 
                     &         &             &    320.1        &  3.1   \\ 
                     &         &             &    346.5        &   4.6  \\ 
                     &         &             &    422.1        &   83.7  \\ 
                     &         &             &    569.3        &   4.8  \\ 
                     &         &             &    578.6        &  7.3   \\ 
                     &         &             &  657.5            &  60.6\\ 
                     &         &             &  \textbf{960.7}      &  99.3\\
                     &         &             &  1245.5            &  2.8\\  
$^{203}$Bi($3n$)  &  11.76 h  &    9/2$^{-}$  &  \textbf{820.2}        & 30.0 \\
                     &         &             &    825.2        &  14.8   \\ 
                     &         &             &    896.9        &  13.2   \\                    
$^{199}$Tl($\alpha$$3n$)  &  7.42 h    &    1/2$^{+}$  &  158.4   &  5.0\\
                              &            &       &  \textbf{208.2 }  &  12.3 \\
                               &            &       &  247.3  &  9.3 \\
                             &            &         &  455.5  &  12.4\\
$^{200}$Tl($\alpha$$2n$)  &   26.1 h    &    2$^{-}$  &  \textbf{367.9}   &  87.0\\
                     &         &             &  579.3            &  13.7\\
                     &         &             &  828.3            &  10.8\\
                     &         &             &  1205.8            &  30.0\\                       
\end{tabular}}
\end{ruledtabular}
\end{table}

\begin{figure}[!t]
\centering
\includegraphics[width=0.35\textwidth]{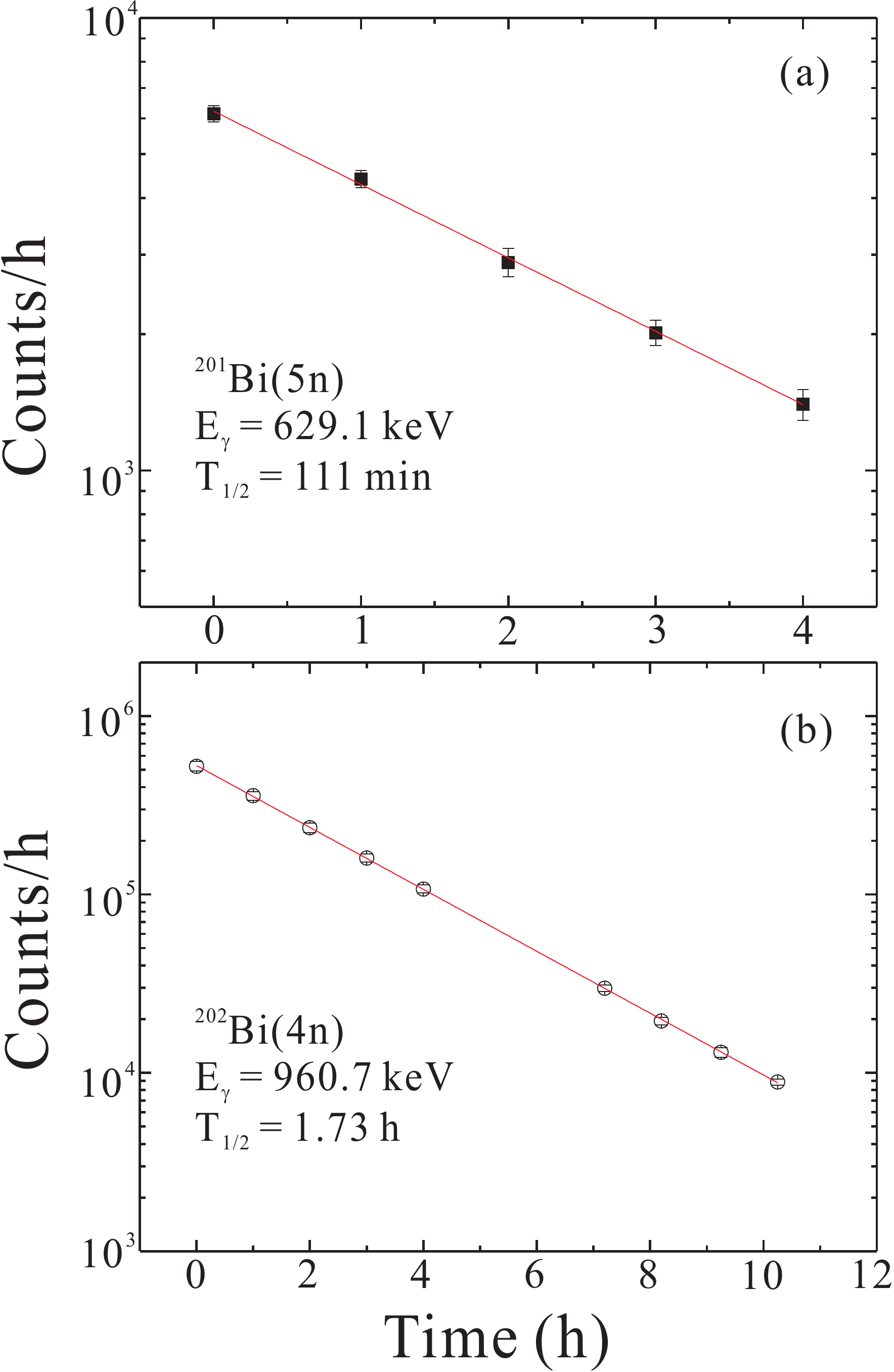}
\caption{\label{fig04} (Color online) Radioactive decay curves for the $ ^{201} $Bi (a) and $ ^{202} $Bi (b) nuclei formed in the $ ^{9} $Be + $ ^{197} $Au reaction by using the 629.1- and 960.7-keV $ \gamma $ rays, respectively.}
\end{figure}

\begin{figure}[t]
\centering
\includegraphics[width=0.35\textwidth]{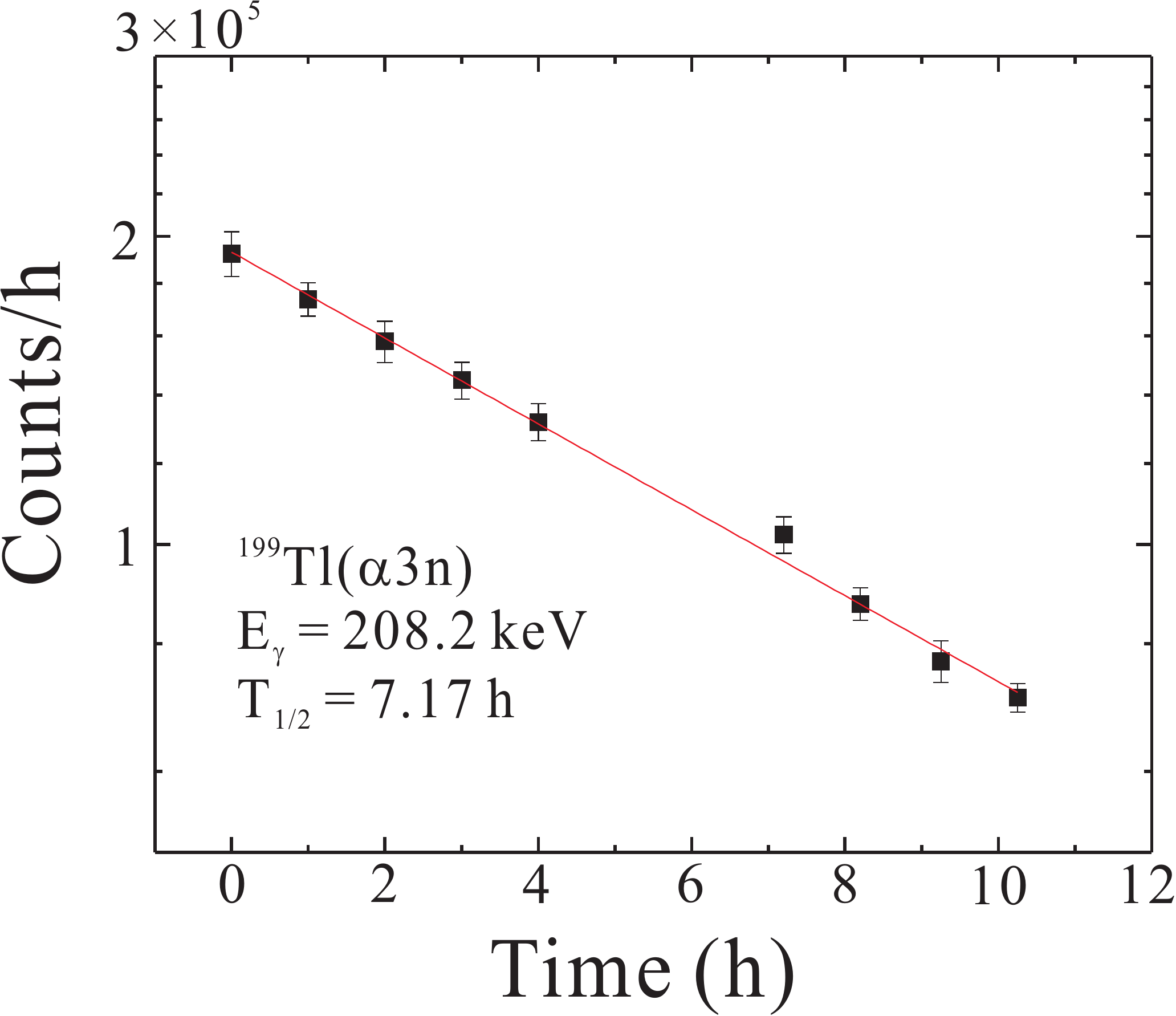}
\caption{\label{fig05} (Color online) Radioactive decay curve for the $ ^{199} $Tl nucleus formed in the $ ^{9} $Be + $ ^{197} $Au reaction by using the 208.2-keV $ \gamma $ ray.}
\end{figure}

The excited compound nuclei formed in the fusion of $ ^{9} $Be + $ ^{197} $Au decay most favorably by neutron evaporations, and $3n$, $4n$, and $5n$ channels from CF are all observed in the present experiment. In addition, the $ \alpha $$2n$ and $ \alpha $$3n$ products were also identified. These nuclei were identified not only by their characteristic $ \gamma $-ray energies, but also by their half-lives and branching ratios, as listed in Table \ref{table01}. Figure \ref{fig02} (a) present the offline $ \gamma $-ray spectrum for the $ ^{9} $Be + $ ^{197} $Au system from the first target (see Fig. \ref{fig01}), measured at 10.5 h after the end of the activation. One could see that the $ \gamma $ rays from the reaction products of $ ^{202} $Bi, $ ^{198} $Au, $ ^{199} $Tl, and $ ^{200} $Tl could be clearly identified. The relatively weak intense $ \gamma $ rays from $ ^{202} $Bi could be identified from the spectrum measured 20 min after the activation, as shown in Fig. \ref{fig02}  (b). The $ \gamma $ rays from $ ^{201} $Bi could also be seen from the spectrum. To justify the identification of $ ^{203} $Bi, we present in Fig. \ref{fig03} the offline $\gamma $-ray spectra measured at 10.5 h after the end of the activation from the first, third, and fifth targets, with a measuring time of 1 h. One could see that the peaks of characteristic $ \gamma $ rays from $ ^{203} $Bi decrease as the beam energy decreases when passing through the target assembly. The radioactive decay curves obtained for CF residues of $ ^{201} $Bi (629.1 keV line) and $ ^{202} $Bi (960.7 keV line) are shown in Figs. \ref{fig04} (a) and (b), respectively. The curve for ICF residue of $ ^{199} $Tl (208.2 keV line) is shown in Fig. \ref{fig05}. The half-lives extracted from our measurements are in agreement with the data in the literature \cite{KONDEV2007365,ZHU2008699,SINGH200779}.

\section{Data reduction and results}\label{sec03}

\begin{table*}[tbp]
\caption{\label{table02} Measured cross sections for the residues formed through the $ ^{9} $Be + $ ^{197} $Au reaction. The effective beam energies in the first column were calculated from the weighted averages of beam energy at the Au foil center and that at the previous one. See text for details.}
\begin{ruledtabular}
\setlength{\tabcolsep}{0.4mm}{
\begin{tabular}{ccccccc}
{E$_{lab}$(MeV)} &
{Target label} &
{$^{201}$Bi(mb)} &
{$^{202}$Bi(mb)} &
{$^{203}$Bi(mb)} &
{$^{199}$Tl(mb)}&
{$^{200}$Tl(mb)}\\[1mm] \hline
     &1st & 74.13$\pm$5.22   & 244.05$\pm$18.39 &                & 121.52 $\pm$16.58  &  64.47$\pm$4.89   \\  
48.9 &2nd & 71.98$\pm$6.56   & 338.91$\pm$37.24 & 11.75$\pm$2.37  & 169.70 $\pm$21.59  &  91.03$\pm$6.92   \\  
47.9 &3rd & 38.95$\pm$4.65   & 354.25$\pm$34.38 & 13.14$\pm$1.59  & 147.87 $\pm$12.77  &  96.68$\pm$6.36   \\  
46.9 &4th & 13.17$\pm$2.70   & 332.55$\pm$28.74 & 17.34$\pm$1.86  & 123.55 $\pm$10.87  & 104.74$\pm$7.52   \\  
45.7 &5th &                  & 270.81$\pm$30.87 & 21.61$\pm$2.36  & 108.86 $\pm$18.61  & 107.67$\pm$7.63   \\     
\end{tabular}}
\end{ruledtabular}
\end{table*}

\begin{table}[b]
\caption{\label{table03} Measured cross section of complete fusion (CF) after correction with the ratio \textit{R} (see text for definition) obtained from PACE, and deduced ICF/TF ratios for the $ ^{9} $Be + $ ^{197} $Au system.}
\begin{ruledtabular}
\setlength{\tabcolsep}{4mm}{
\begin{tabular}{cccc}
{E$_{lab}$(MeV)} &
\textit{R} &
{$ \sigma $ $ _{CF}^{exp}$ (mb)}&
{ICF/TF}\\[1mm] \hline
48.9 &	0.957    & 	 	441.63$\pm$39.59    & 	 	0.37      \\  
47.9 &	0.963    & 	 	421.95$\pm$36.06    & 	 	0.37      \\  
46.9 &	0.963    & 	 	377.01$\pm$30.04    & 	 	0.38     \\   
45.7 &	0.961    & 	 	304.29$\pm$32.21    & 	 	0.42     \\        
\end{tabular}}
\end{ruledtabular}
\begin{flushleft}
\end{flushleft}
\end{table}

The preliminary experimental cross sections of products formed in the $ ^{9} $Be + $ ^{197} $Au reaction on each of the targets were extracted using the half-lives, prominent $ \gamma $-ray energies of decay, and intensities as well the formula described in Ref.~\cite{zhang2014complete}. The results are given in Table \ref{table02}. One could see from the preliminary results that the yields of the first target show about one-third less, when comparing with the systematic trend of the other four targets (e.g., cross sections of Bi nuclei). This is because no catcher was available for the studied reaction system, and part of the produced residues penetrated to the subsequent target during the online activation. Therefore, cautious calculation is required in order to associate effective beam energies to the reaction product yields for each target unambiguously. To further estimate the percentage of loss in each target, we employed PACE4 \cite{PhysRevC.21.230} to calculate the energies and angular distributions of residues at each target. These results were then used as the inputs of SRIM \cite{ZIEGLER20101818} calculation, assuming those residues were positioned uniformly among the Au foils upon produced. The calculation showed that in all five Au foils about 30\% of the fusion evaporation residues lost to the subsequent substance. The effective beam energies on the last four Au foils were calculated following the weighted average:
\begin{linenomath*}
\begin{equation}
\label{eq01} {E_{eff}} = {\frac{{\sum_{i}E_{i}P(E_{i})\sigma(E_{i})}}{{\sum_{i}P(E_{i})\sigma(E_{i})}}},
\end{equation}
\end{linenomath*}
where $ P(E_{i}) $ represents the probability that the reaction residue inside the target is associated with beam energy $ E_{i} $, and $ \sigma(E_{i}) $ refers to the corresponding fusion cross section at $ E_{i} $, calculated by the PACE4 code. The deduced energy values are presented in the first column of Table \ref{table02}. It should be pointed out that the inaccuracy of SRIM calculation does not affect the deduced effective energies too much. For example, a 20\% difference of calculated residue loss in the Au foil gives the deduced energy difference of about 0.2 MeV. Fisichella \textit{et al.} \cite{PhysRevC.92.064611} have pointed out the possibility of misinterpretations of a derived excitation function resulted from the ambiguities of derived beam energies, typically in the exponential region of cross section below the barrier. In our work, the beam energies are above the barrier energies, and the changes of fusion cross sections are not dramatic. Therefore, the inaccuracy of the adopted beam energies will affect little on the physics discussions of this work. The errors of the deduced cross sections are the combination of statistical error and errors due to target thickness ($ \approx $3\%), beam current ($ \approx $3\%), and detector efficiency ($ \approx $3\%).

The CF cross sections were deduced through dividing the cumulative measured ($ \sigma $$ _{3n+4n+5n} ^{exp} $) cross sections by the ratio $R$, which gives the missing ER contribution, if any. Here the ratio $R$ refers to $\Sigma_{x}$$\sigma_{xn}^{PACE4}$/$\sigma_{fus}^{PACE4}$ , where $x$ = 3, 4, 5. The dominant ICF channels are found to be $ \alpha 2n$ ($ ^{200} $Tl), $ \alpha 3n$ ($ ^{199} $Tl). The possible $ \alpha 1n$ ($ ^{201} $Tl) channel was not found in this work. Note that the Tl isotopes can be not only from ICF but the sum of ICF plus a possible contribution from CF with $ \alpha $ evaporation. However, the PACE4 code predicts that the CF compound nuclei formed in the $ ^{9} $Be + $ ^{197} $Au system at the measured energy range decay overwhelmingly by neutron evaporation, and the total evaporation of $\alpha$ channels is only about 1\%. Therefore, the sum of the $ \alpha 2n$ and $ \alpha 3n$ channels was used in the present work to estimate the ICF cross sections. The TF cross sections were deduced by adding the corresponding CF and ICF cross sections. The values of ratio $R$, deduced CF cross sections as well as ICF/TF ratios are listed in Table \ref{table03}. 

\section{Discussion}\label{sec04}

\begin{figure}[!t]
\centering
\includegraphics[width=0.4\textwidth]{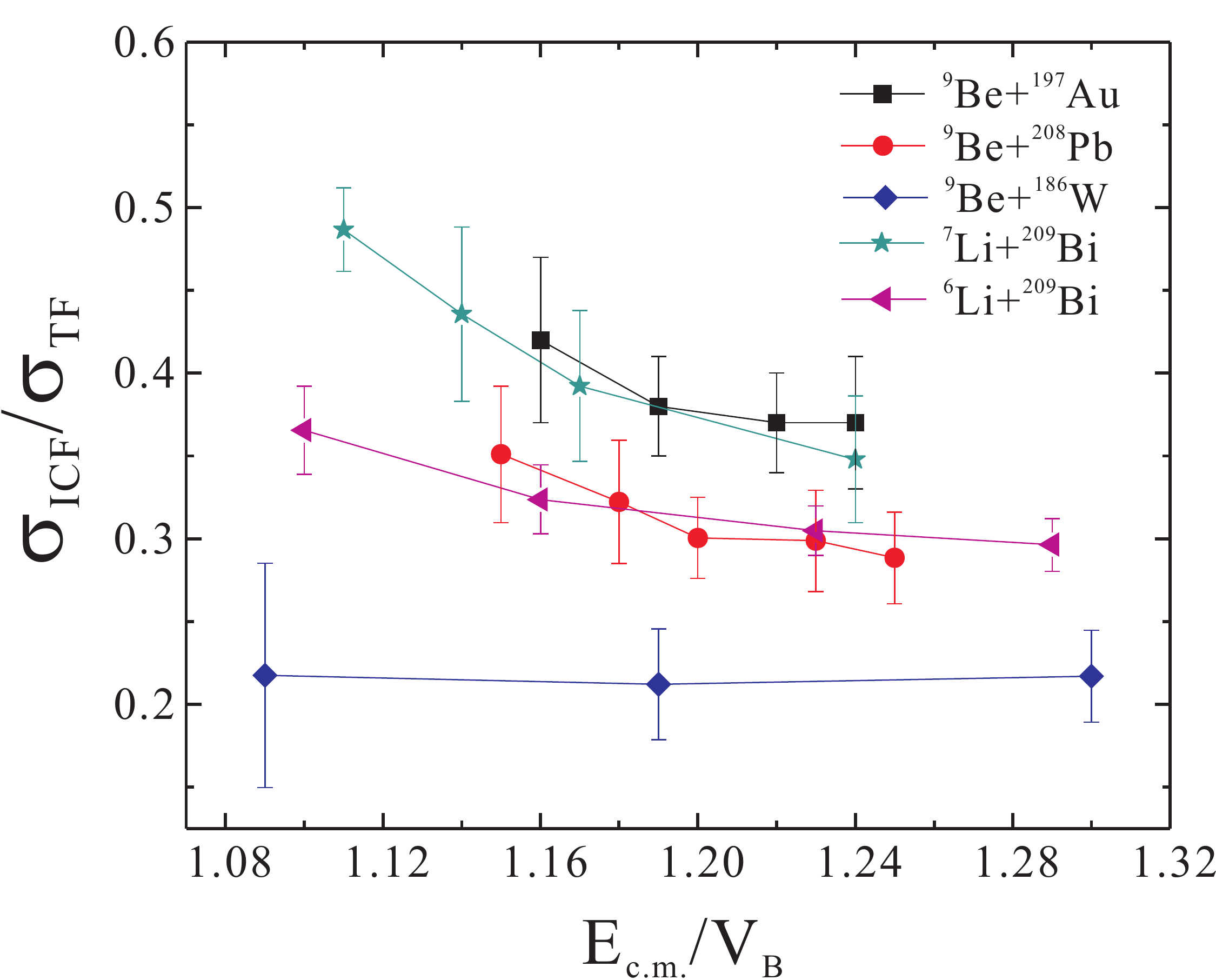}
\caption{\label{fig06} (Color online) Model-independent of ratios of incomplete fusion (ICF) and total fusion (TF) cross sections measured at above barrier energies in the $^{9} $Be + $ ^{197} $Au (present work),$ ^{208} $Pb \cite{dasgupta2004effect,dasgupta1999fusion}, $ ^{186} $W \cite{fang2013fusion}, and $^{6,7} $Li + $ ^{209} $Bi \cite{dasgupta2004effect} reaction systems, as a function of the energy relative to the Coulomb barrier.}
\end{figure}

The ICF probabilities, defined as the ratios between ICF and IF cross sections, for the $^{9} $Be + $ ^{197} $Au system at above barrier energies are around 0.4, slightly higher than the other systems with $^{9} $Be projectiles reported in the literature for the $^{144} $Sm~\cite{gomes2006disentangling,GPC06}, $ ^{169}$Tm~\cite{PhysRevC.91.014608}, $ ^{181}$Ta~\cite{zhang2014complete}, $ ^{186} $W~\cite{fang2013fusion}, $ ^{187} $Re~\cite{zhang2014complete}, and $ ^{208} $Pb~\cite{dasgupta2004effect,dasgupta1999fusion} targets. In this work, we chose  $^{9} $Be + $ ^{208} $Pb \cite{dasgupta2004effect,dasgupta1999fusion}, $ ^{186} $W \cite{fang2013fusion}, and $^{6,7} $Li + $ ^{209} $Bi \cite{dasgupta2004effect} reaction systems for comparison. The ratios are shown in Fig. \ref{fig06}, as a function of the quantity $E_{c.m.}$/$V _{B} $, where $E_{c.m.}$ refers to energy in the center of mass frame, and $V _{B} $ is the height of the Coulomb barrier. It should be pointed out that the ratios in the figure are model-independent and require the measurement of both ICF and CF, which is available up to now for only less than ten reaction systems with $^{9} $Be projectiles.

\begin{figure}[b]
\centering
\includegraphics[width=0.43\textwidth]{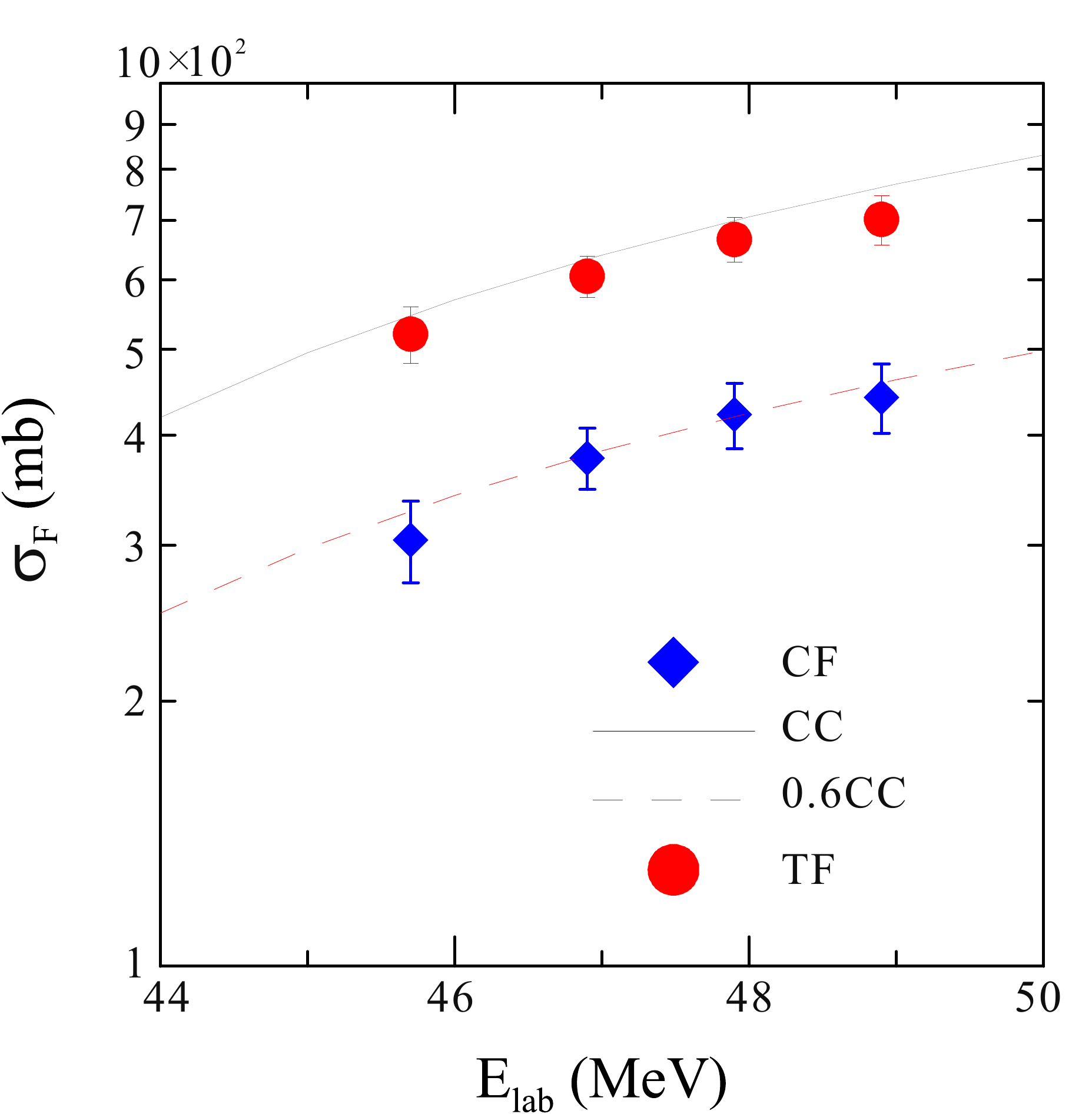}
\caption{\label{fig07} (Color online) Comparison of experimental complete fusion (CF) and total fusion (TF) functions for the $^{9} $Be + $ ^{197} $Au system with coupled-channel calculations that do not include the breakup channel. See text for details.}
\end{figure} 

\begin{figure}[t]
\centering
\includegraphics[width=0.43\textwidth]{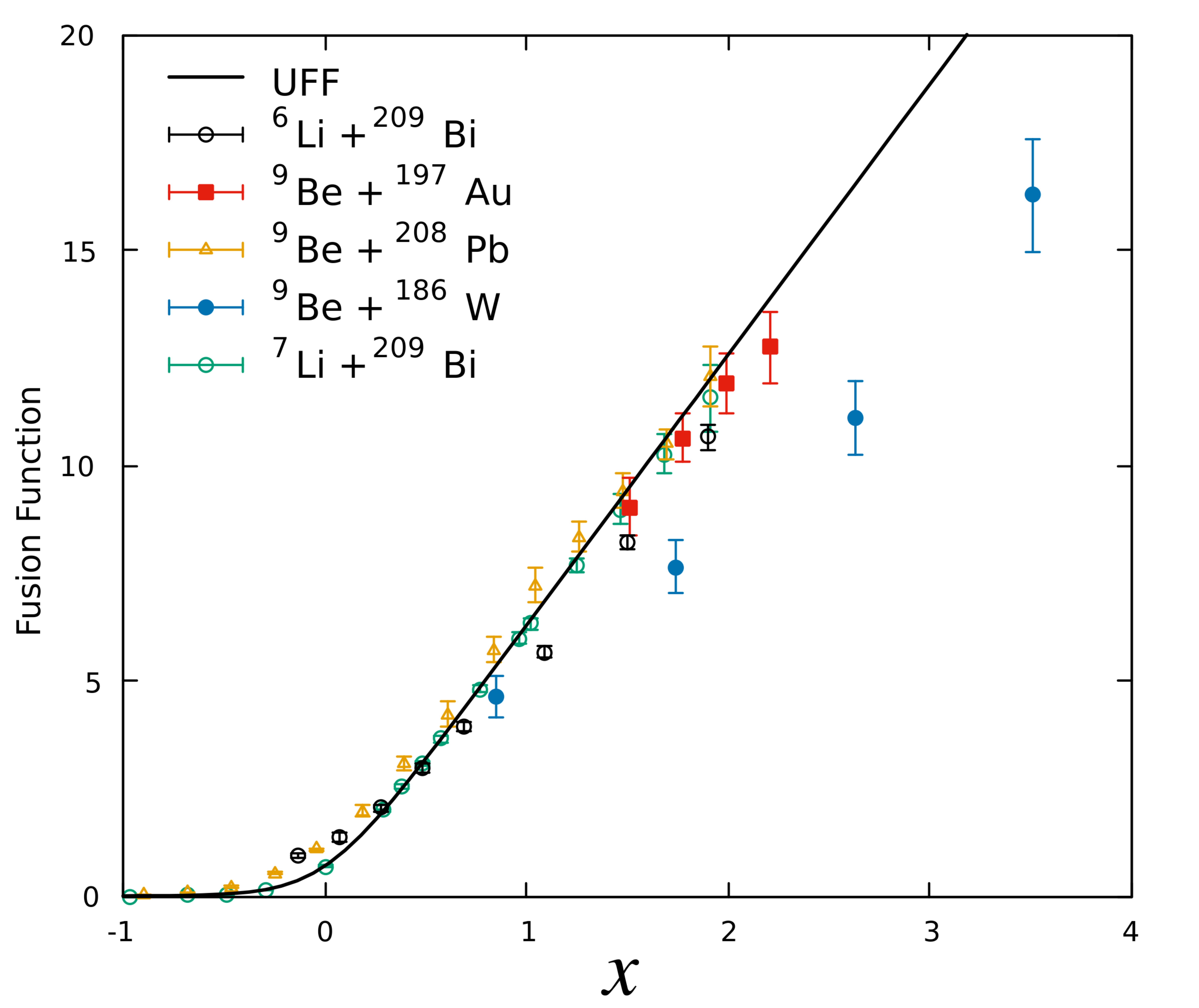}
\caption{\label{fig08} (Color online) Comparasion between universal fusion function (UFF) and renormalized experimental fusion function for total fusion (TF). The data for $^{9}\textrm{Be}+^{208}\textrm{Pb}$, $^{6,7}\textrm{Li}+^{209}\textrm{Bi}$, and $^{9}\textrm{Be}+^{186}\textrm{W}$ systems were obtained from Refs.~\cite{dasgupta2004effect,dasgupta1999fusion,fang2013fusion}.}
\end{figure} 

\begin{figure}[t]
\centering
\includegraphics[width=0.43\textwidth]{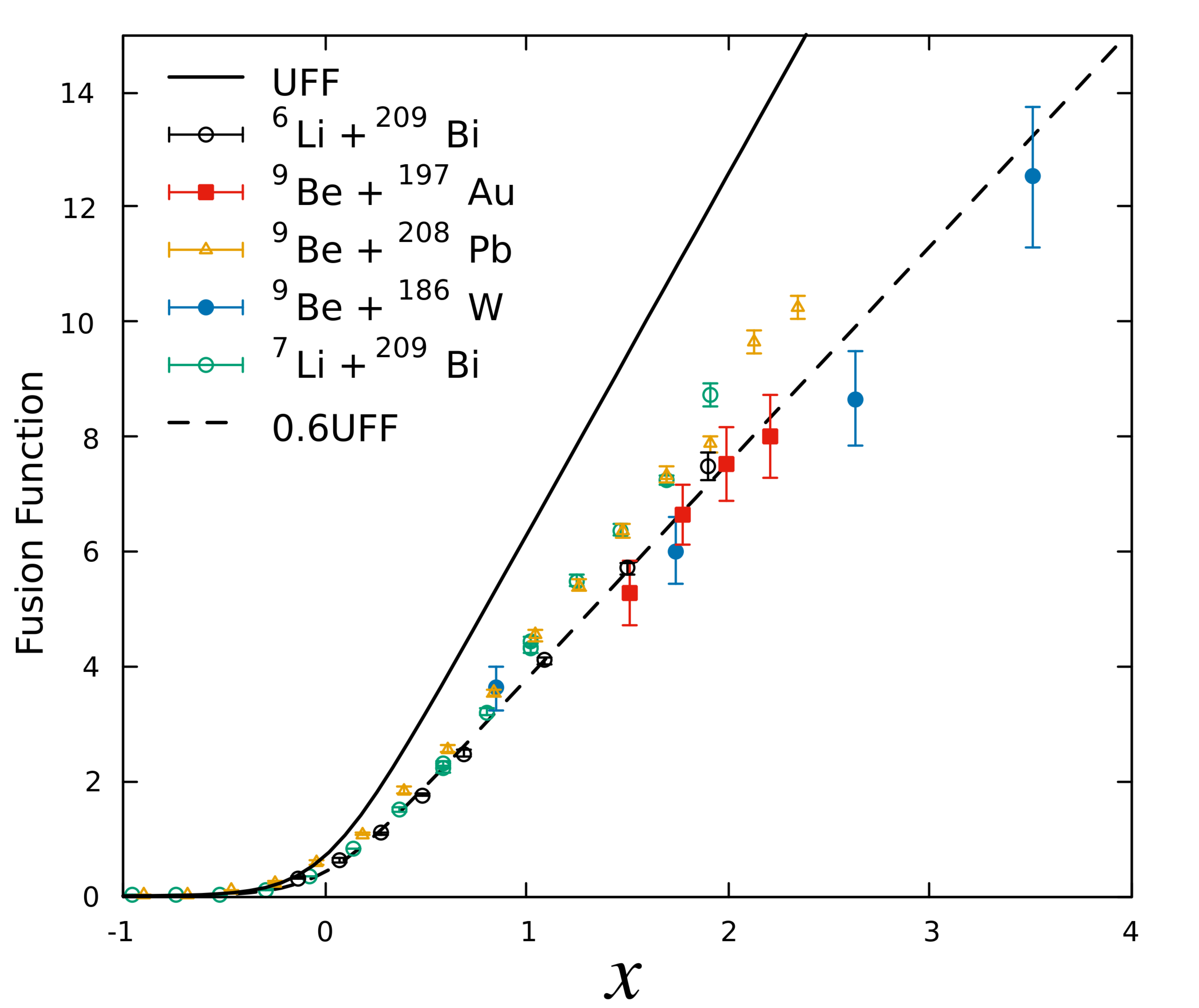}
\caption{\label{fig09} (Color online) Comparasion between universal fusion function (UFF) and renormalized experimental fusion function for complete fusion (CF). The dashed line is the UFF multiplied by 0.6. The data for $^{9}\textrm{Be} \; + \; ^{208}\textrm{Pb}$, $^{7}\textrm{Li} \; + \; ^{209}\textrm{Bi}$, and $^{9}\textrm{Be} \; + \; ^{186}\textrm{W}$ systems were obtained from  Refs. \cite{dasgupta2004effect,dasgupta1999fusion,fang2013fusion}. }
\end{figure}

The CC calculation was first performed to study the effect of coupling on fusion cross sections. In the calculation, we chose a parameter-free S\~ao Paulo potential (SPP) \cite{chamon1997lc,chamon2002lc} as the real part of the optical potential. This is a double-folding potential with systematic matter density that accounts for Pauli non-locality in the exchange of nucleons. The barrier parameters predicted by the SPP are $R_{B}$ = 11.35 fm, $V_{B}$ = 37.50 Mev, and $\hbar \omega$ = 4.44 MeV. The imaginary part of the potential used was a Woods-Saxon form with parameters ($W$= -50 MeV, $r_w$= 1.06 fm, and $a_w$ = 0.2 fm for the depth, reduced radius, and diffuseness, respectively) to ensure that the absorption occurred only when the barrier was tunneled or overcome. The calculations were performed using the FRESCO code \cite{thompson1988ij}. The ground state and first five excited states of $^{197}\textrm{Au}$ were included in the calculation. The deformation parameter of $^{197}\textrm{Au}$ was taken from Ref. \cite{wallmeroth1989nuclear}. No coupling associated to $^{9}\textrm{Be}$ was included in the calculation. The resulting effects can be assumed as coming from the whole dynamic effect of the $^{9}\textrm{Be}$ breakup on the CF cross section. Figure \ref{fig07} shows the results of calculations, in comparison with the experimental CF and TF data. From the figure, it can be seen that the cross sections obtained from the CC calculation are higher than the experimental CF values but in agreement with the TF values, indicating a suppression (about 40\%) in measured CF cross sections at above-barrier energies.

In order to make a comparison with other reaction systems, we further employed the UFF methodology \cite{Canto_2008,CANTO200951} in the analysis. The function follows a reduction procedure that withdraws the dependency on the statics effects of the weakly bound nucleons and reveals the relevance of channel couplings on the fusion. To obtain the reduced cross section and collision energy, one applies the reduction procedure in the form
\begin{linenomath*}
	\begin{equation}
	E \rightarrow x = \frac{E_{c.m.} - V_{B}}{\hbar \omega}, \; \; \sigma_{F} \rightarrow F(x) = \frac{2E_{c.m.}}{\hbar\omega R^{2}_{B}}\sigma_{F},
	\label{eq:ff}
	\end{equation}
\end{linenomath*}
where $\sigma_{F}$ is the fusion cross section, $R_{B}$ and $\hbar \omega$ are the radius and curvature of the Coulomb barrier, respectively. The reduction procedure was inspired by Wong's formula  \cite{wong1973cy}. Approximating the Coulomb barrier to a parabola and assuming that it is independent of the angular momentum, an analytical expression is obtained for the fusion cross section:
\begin{linenomath*}
	\begin{equation}
		\sigma_{F}^{W} = \frac{\hbar \omega R^{2}_{B}}{2E_{c.m.}} ln \left[ 1 + exp \left( \frac{2\pi(E-V_{B})}{\hbar \omega}\right) \right].
	\label{eq:wong}
	\end{equation}	
\end{linenomath*}
This formula was extensively used in the past to describe the fusion cross section by varying the three parameters ($V_{B}$ , $R_{B}$, and $\hbar \omega$). Applying the reduction procedure given in formula (\ref{eq:ff}), the UFF could be written as
\begin{linenomath*}
	\begin{equation}
		F_{0}(x) = 	ln [1 + exp(2\pi x)],
	\label{eq:uff}	
	\end{equation}
\end{linenomath*}
which is system independent. As pointed out by Canto \textit{et al.} in Refs. \cite{Canto_2008,CANTO200951}, the reduction method has two shortcomings. One is that Wong's formula is not accurate in describing the fusion cross section of light systems, as in those systems the Coulomb barrier could not be directly approximated as a parabola. The other is that the effect of the breakup process on fusion could not be inferred from the comparisons of $F_{exp}(x)$ with UFF. To account for these shortcomings, Canto \textit{et al}. \cite{Canto_2008,CANTO200951} introduced a fusion function renormalized by CC calculations. This function is defined as $\bar{F}_{exp} = F_{exp}\frac{\sigma_{F}^{W}}{\sigma_{F}^{CC}}$, where $\sigma_{F}^{CC}$ is the cross section obtained by CC calculations, and $F_{exp}$ is the reduced experimental fusion cross section. It is important to mention that in a perfect situation where all couplings effects were considered in the CC calculation, the renormalized fusion function ($\bar{F}_{exp}$) is identical to UFF.
		
The renormalized experimental fusion function for TF of the $^{9} $Be + $ ^{197} $Au system is shown in Fig. \ref{fig08}, in comparison with the UFF. The corresponding data of $^{9} $Be + $ ^{208} $Pb, $ ^{186} $W, and $^{6,7} $Li + $ ^{209} $Bi reaction systems are also presented for comparison. We choose a linear scale in the figure as it is more suitable for the analysis at above barrier reaction energies ($x = 0$ in the figure corresponds to the fusion barrier). The barrier parameters and the coupling scheme used in CC calculations for the systems not measured in the present work are given in Refs.~\cite{CANTO200951,fang2013fusion}. One can see from Fig. \ref{fig08} that the $^{9} $Be + $ ^{197} $Au, $ ^{208} $Pb, and $^{6, 7} $Li + $ ^{209}$Bi systems are in good agreement with the UFF, showing the consistency of this reduction method. The $^{9} $Be + $ ^{186} $W system shows a suppression of about 25 \%, compared to the other systems. This is probably caused by the large portion of the one-neutron stripping process, which can be evidenced by the vanish of suppression when adding the one-neutron stripping at TF \cite{fang2013fusion}.

Figure \ref{fig09} is similar to Fig. \ref{fig08}, but it shows the CF function. The difference between the points and the UFF curve is the observed effects of the breakup of projectiles plus transfer channels on the CF for all systems. It is interesting to mention that the $^{9} $Be + $ ^{197} $Au system shows a suppression of about of 40\% compared to the UFF. This is a bit higher than the $^{9} $Be + $ ^{208} $Pb \cite{CANTO200951}, $ ^{181} $Ta \cite{zhang2014complete}, $ ^{169} $Tm and $ ^{187} $Re \cite{PhysRevC.91.014608}, and $^{6,7} $Li + $ ^{209} $Bi systems \cite{CANTO200951} that show a suppression of about of 30$\sim $ 35\%. The CF suppression factor of the $^{9} $Be + $ ^{186} $W system deserves further investigation, as the renormalized fusion function for TF is not consistent with the UFF. The results show the success of the reduction method in analyzing the fusion reaction of different reaction systems, which revealed a systematic hindrance of the CF at energies above the Coulomb barrier for heavy systems involving stable weakly bound projectiles.

To finish the discussion on the hindrance of the TF and CF cross section at energies above the Coulomb barrier, we would like to emphasize that the separation of CF from TF is not an easy task, from both the theoretical and experimental points of view. For this reason, to contribute with more experimental data including weakly bound (stable or radioactive) nuclei, that allows to arrive at a definite conclusion about the effect of the breakup channel on CF and TF, is very important. As already mentioned, the effect of the breakup plus transfer channels on the TF was not very clear for the systems involving the $^9$Be projectile. From the present results one can conclude that the CF is hindered at energies above the barrier, and the TF is not affected by the breakup channel at this energy regime. Concerning the mass dependence of the hindrance of the CF above the barrier, the amount of the experimental data is still not enough. More precise experimental TF and CF data for medium-mass systems are mandatory to arrive at a definite conclusion. In addition, studies \cite{CSB19,PhysRevLett.122.042503,PhysRevC.97.021601} have also shown that the main reason for the CF hindrance is the transfer of clusters, rather than the breakup followed by the absorption of the fragments. The conclusion is currently controversial. To help to clarify this it is very important to develop a full quantum mechanical method that could derive CF, ICF, and TF.
	
\section{Summary}\label{sec05}	
In this work, we report the measurement of complete and incomplete fusion cross sections for the $^{9} $Be + $ ^{197} $Au system at above barrier energies, using the online activation and offline $\gamma$-ray spectroscopy method. Comparison of data with coupled channel calculations, that do not take into account the breakup and transfer channels, shows a CF suppression of about 40\% at above barrier energies. The TF excitation function is in agreement with the theoretical prediction. In addition, we compared the behaviors of the TF and CF functions with those in the $^{9} $Be + $ ^{208} $Pb, $ ^{186} $W, and $^{6,7} $Li + $ ^{209} $Bi reaction systems, by employing the universal fusion function methodology. It revealed that for the TF only the $^{9} $Be + $ ^{186} $W system is not consistent with the universal fusion function, while in the other four systems the $^{9} $Be + $ ^{197} $Au shows a relatively larger CF suppression compared to the $^{9} $Be + $ ^{208} $Pb and $^{6,7} $Li + $ ^{209} $Bi systems. Once the understanding of the effect of the breakup plus transfer channels at energies above the barrier is clear, it is important to have more experimental data at energies below the Coulomb barrier, to try to understand the details of the reaction mechanism at this energy regime.

\section*{ACKNOWLEDGMENTS}
The work at Institute of Modern Physics, CAS was supported by the National Key R\&D Program of China (Contract No. 2018YFA0404402), the Youth Innovation Promotion Association of Chinese Academy of Sciences (Grant 2019407), the Fundamental Research Funds of Chinese Academy of Sciences (Grant QYZDJ-SSW-SLH041), and the  National Natural Science Foundation of China (Grant Nos.11305221, 11575255, U1732139). The work at Universidade Federal Fluminense was supported by CNPq, FAPERJ, and CAPES and from INCT-FNA (Instituto Nacional de Ci\^{e}ncia e Tecnologia- F\'{i}sica Nuclear e Aplica\c{c}\~{o}es), research project 464898/2014-5.

\bibliography{mybibfile}

\end{document}